\documentclass[fleqn,useAMS,usenatbib]{mnras}

\usepackage{newtxtext}

\usepackage[T1]{fontenc}
\usepackage{ae,aecompl}

\usepackage{color}
\usepackage[english]{babel}
\usepackage{graphicx}
\usepackage{amsmath}
\usepackage{amssymb}
\usepackage{bm}
\usepackage[utf8]{inputenc}
\usepackage{ulem}

\begin{document}

\newcommand{\dd}{\mathrm{d}}
\newcommand{\rhob}{\rho_\text{b}}
\newcommand{\Tb}{T_\text{b}}
\newcommand{\kB}{\mbox{$k_\text{B}$}}
\newcommand{\rate}{\mbox{erg cm$^{-3}$ s$^{-1}$}}
\newcommand{\gcc}{\mbox{g~cm$^{-3}$}}
\newcommand{\xg}{x_\text{g}}
\newcommand{\rg}{r_\text{g}}
\newcommand{\gs}{g_\text{s}}
\newcommand{\me}{m_\text{e}}
\newcommand{\Ts}{T_\text{s}}
\newcommand{\hatn}{\hat{\bm{n}}}
\newcommand{\hatk}{\hat{\bm{k}}}
\newcommand{\SB}{\sigma_{\rm SB}}
\newcommand{\Omegas}{\Omega_\text{s}}
\newcommand{\Teff}{T_\text{eff}}
\newcommand{\Bpole}{B_\text{pole}}
\newcommand{\Beff}{B_\text{eff}}
\newcommand{\scc}{s_\text{c}}
\newcommand{\shh}{s_\text{h}}
\newcommand{\pcc}{p_\text{c}}
\newcommand{\phh}{p_\text{h}}

\newcommand{\Reff}{R_\mathrm{eff}^\infty}

\newcommand{\sSB}{\sigma_\mathrm{SB}}

\newcommand{\Ti}{\widetilde{T}}
\newcommand{\Lnu}{L_\nu^\infty}
\newcommand{\Ls}{L_\gamma^\infty}

\newcommand{\msun}{{\rm M} \odot}
\newcommand{\xmm}{\textit{XMM-Newton}}
\newcommand{\chan}{\textit{Chandra}}
\newcommand{\fermi}{\textit{Fermi}}


\title[Two-BB portraits of radiation from magnetized neutron stars]{Two-blackbody portraits of radiation from magnetized neutron stars}

\author[D. G. Yakovlev]{ 
{D. G.    Yakovlev$^{1}$\thanks{E-mail:
yak.astro@mail.ioffe.ru}}
\\
$^{1}$ Ioffe Institute, Politekhnicheskaya 26, St~Petersburg 194021,
Russia\\
%
}

\date{Accepted . Received ; in original form}
\pagerange{\pageref{firstpage}--\pageref{lastpage}} \pubyear{2010}

\maketitle \label{firstpage}

\begin{abstract} 

We study a simple model describing thermal
radiation spectra from magnetized neutron stars. The model assumes that
a star is nearly spherical and isothermal inside and possesses 
dipole magnetic fields ($B \lesssim 10^{14}$ G) near the surface, which 
make the surface temperature distribution non--uniform.  
We assume further that any surface element emits a blackbody (BB)
spectrum with a local effective temperature.
We show that such thermal spectra (including phase--resolved) are accurately 
approximated by simple equivalent two--BB (2BB) models. We introduce and study 
phase--space maps of 2BB parameters and show that these maps 
can be useful for interpreting neutron star observations, in which 2BB spectral
fits have been done. 
\end{abstract}


\section{Introduction}
\label{s:introduc}

Thermal radiation of neutron stars is formed in very thin surface layers 
(of thickness from few centimeters to a few millimeters, as
reviewed, e.g. by \citealt{Potekhin14}). This radiation contains important 
information on neutron star parameters (such as mass, raduis,
chemical composition and magnetic field in surface layers).
The information can be inferred by interpreting observations
of neutron stars with theoretical atmosphere models or models
of radiating condensed surface.
These interpretations can also be useful
for exploring superdense matter in neutron star interiors
(see, e.g., \citealt{Potekhin14,PPP2015a} and references therein).

We study thermal radiation, 
which emerges from stellar interiors of middle-aged isolated 
neutron star.
We neglect non-thermal radiation component
(of magnetospheric origin) and the effects of rapid rotation. 
We will mainly consider simplest  
emission models of neutron stars 
which possess purely dipole magnetic fields in the surface layers
with typical field strengths $B \lesssim 10^{14}$~G. 

The magnetic field can make
the surface temperature distribution anisotropic and 
modify the radiation spectrum  
(e.g. \citealt{PPP2015a,PPP2015b,grandis21}). The anisotropic surface temperature 
distribution in a middle-aged 
neutron star is usually created in the so-called heat blanketing
envelope (e.g. \citealt{BPY21}), which thermally insulates a warm and nearly isothermal interior of the
star from much cooler surface. This heat blanket is typically much
thicker than the atmosphere.  We will 
take the surface temperature distrbutions from the
models of heat blankets (e.g.
\citealt{PYCG03,HoPC08,PPP2015b}, and references therein). 

The magnetic field effects on thermal emission from the atmosphere 
are complicated (see, e.g. \citealt{2002ZAVPAV,PPP2015a,gonzalezea2016}). 
The radiation can be strongly
polarized, contain specific spectral and angular features; it can be greatly variable over the surface. The radiation flux
detected by a distant observer must be calculated by integration over the visible
part of the surface, taking into account gravitational redshift and light bending 
of photons.

The progress in developing the magnetized atmospheric models of neutron stars has been reviewed, for 
instance, by \citet{Potekhin14,PPP2015a}. 
Such models can be separated into those
with a pure radial magnetic field constant on the surface and those with more realistic magnetic
fields varying over the surface. 
The first models are easier to construct but seem less realistic; we will not discuss
them. We focus on the atmosphere models with the dipole field configuration which are not
numerous (e.g. \citealt{HoPC08,ZKSP21}). We consider a simplified version of
such models, in which any local element on the stellar surface
(with non-uniform temperature distribution) emits like a blackbody (BB) with
its own local surface temperature $\Ts$, meaning an overall multi--BB thermal radiation. The model
was put forward long ago \citep{Greenstein} and analyzed in the literature 
(e.g. \citealt{page95,page96,PY01,PYCG03,geppert2006,zane-turolla2006}). Our main goal will be to study its
approximation by a two-BB (2BB) model, a widely used
tool for investigating neutron-star radiation.

In Sect.\ \ref{s:model} we outline a general multi--BB radiation model  
for a neutron star with isothermal interiors and dipole surface
magnetic field. In Sect.\ \ref{s:2BB} we study spectral fluxes emitted
from such a star. We show that these fluxes can accurately be
fitted by 2BB models and analyse the dependence of fit parameters on
the mean effective temperature of the star, magnetic field strength, and
stellar compactness. Also, we derive  accurate 2BB fits for any observation
direction.
 We summarize the results and conclude in Sect.\ \ref{s:discuss}.

\section{Theoretical multi--BB spectra}
\label{s:model}

\subsection{Theoretical outline}
\label{s:TheoryOutline}

Let us outline standard calculation of radiation
from a spherical neutron star with {\it any} (not necessarily axially symmetric) non-uniform surface temperature
distribution, assuming that
any surface element emits a BB  radiation 
with an effective (local, non-redshifted) surface temperature $\Ts$, resulting in multi--BB emission.
Since neutron stars are compact,
one needs to account for gravitational light bending
and redshift in Schwarzschild space-time. These effects are characterized by the neutron star compactness parameter
\begin{equation}
    \xg=\frac{\rg}{R}, \quad \rg=\frac{2GM}{c^2},
    \label{e:xg}
\end{equation}
where $M$ is a gravitational stellar mass, $R$ is a circumferential radius, 
$\rg$ is the Schwarzschild radius, $G$ is the gravitational
constant and $c$ is the speed of light. We will use physical quantities
either in a local reference frame near the neutron star surface or
in the frame of a distant observer. The latter 
will be often marked by the symbol $\infty$. For instance, $\Ts^\infty=\sqrt{1-\xg}\,\Ts$,
and the `apparent' neutron star radius for a distant observer is $R_\infty=R/\sqrt{1-\xg}$.

The observer detects the radiative spectral flux density $F_\infty$ 
[erg cm$^{-2}$ s$^{-1}$ keV$^{-1}$] at a distance $D \gg R$  per
unit interval of the redshifted photon energy $E \equiv E_\infty$. The flux depends on
$E$ and observation direction (along a unit vector $\bm{k}$).
Let us take a surface element $\dd S$ of the atmosphere and 
specify its position either by ordinary spherical angles
$\vartheta$ and $\varphi$ (with respect to the magnetic axis) or by
the unit vector $\bm{n}$ normal to the surface.
Then $\dd S=R^2  \, \dd \Omega_{\rm s}$, with
$\dd \Omega_{\rm s} =\sin \vartheta \, \dd  \vartheta \, \dd \varphi $ being a 
solid angle element measured from the stellar centre.
The  contribution of this element to  flux density is
(e.g. \citealt{2002Beloborodov}, \citealt{2003Poutanen} and \citealt{Potekhin14})
\begin{equation}
    \dd F_E^\infty = I(E,\bm{k}_0) \,{\cal P}  
    \, \frac{\dd S}{D^2}, \quad {\cal P}=\cos \theta
    \left| \frac{\dd \cos \theta}{\dd \cos \theta_0} \right|,
\label{e:dF}    
\end{equation}
$\bm{k}_{0}$ being the unit vector along the propagation 
of those photons in the atmosphere, which are registered by the observer. 
The directions of $\bm{k}$ and $\bm{k}_0$ differ because of gravitational
line bending. Furthermore, let $\theta$ (or $\theta_0$) be the angle between
$\bm{k}$ (or $\bm{k}_0$) and 
the line from the stellar centre to the observer. For given $\bm{n}$ 
and $\bm{k}$,  
$\bm{k_0}$ is determined by kinematics of light bending.
Next, $I(E,\bm{k}_0)$ is the radiative intensity emitted
in the atmosphere and redshifted for the observer. Finally,
${\cal P}$ corrects the intensity of the observed flux 
due to light bending; it depends on $\theta$, $\theta_0$ and $x_{\rm g}$. 
Any surface element $\dd S$ is vizible by the observer if
$\theta \leq \pi/2$. Neglecting the light bending, one has  
$\theta=\theta_0$ and ${\cal P}= \cos \theta$.

Using equation (\ref{e:dF}) and integrating over $\dd S$, one obtains
\begin{equation}
F_E^{\infty}=\frac{R^2}{D^2}\,H_E^{ \infty}, \quad
H_E^{ \infty}= \int_{\rm viz} I(E,\bm{k}_0) {\cal P} \, \dd \Omega_{\rm s},
\label{e:H}   	
\end{equation}  
where $H^{\infty}_E$ is the effective 
spectral thermal flux (formally independent of $D$ and $R$) in the 
direction of $\bm{k}$. 
The intergration over $\dd \Omega_{\rm s}$ is
carried over the visible part of the stellar surface.
In our case, the gravitational redshift and light bending are decoupled. 
The light bending enlarges the observable emission area. It
allows one to
observe the emission from the opposite side of 
the star. Note that we neglect
photon absorption by intervening matter
along  propagation path, which will be discussed later.

A local radiation intensity $I_0(E_0)$
of any surface element is 
given by the Planck function with a local non-redshifted
surface temperature $\Ts(\bm{n})$ and non-redshifted 
photon energy $E_0\equiv E/\sqrt{1-\xg}$. Redshifting $I_0$ 
and $E_0$, from
equation (\ref{e:H}) we obtain  
\begin{equation}
H_E^{\infty}=\frac{15 \SB }{16 \pi^5 k_{\rm B}^4}\, 
\int_{\rm viz} \dd \Omega_{\rm s}\,
\frac{(1-\xg)^{-1}\, {\cal P}  E^3}{\exp(E/\kB \Ts^\infty)-1} ,
\label{e:H(Einfty)} 
\end{equation} 
where $\SB$ is the Stefan--Boltzmann constant.
Using accurate analytic approximations of ${\cal P}$ derived by
\citet{2020Poutanen}  [his equations (12) and (13)], 
one can easily compute 
$H_E^{\infty}$ for any surface temperature distribution
$\Ts(\bm{n})$ and observation 
direction $\bm{k}$. In addition, one can 
calculate the total thermal bolometric luminosity
of the star, $L_\text{s}$, and introduce an overall non-redshifted
effective surface temperature $\Teff$ according to (e.g. \citealt{PY01})
\begin{equation}
L_\text{s}=\SB R^2 \int_{4 \pi} \dd \Omegas \, \Ts^4( \bm{n} ) 
\equiv 4 \pi \SB R^2 \Teff^4. 
\label{e:Ls}
\end{equation}
Here one
should integrate over the entire stellar surface.  

In addition to the exact effective flux $H_E^{\infty}$ in the direction of $\bm{k}$, which 
is given by equation (\ref{e:H(Einfty)}), one can
introduce the flux $H_E^{ \rm av \infty}$, averaged over
orientations of $\bm{k}$ at a fixed distance $D$,
\begin{equation}
H_E^{ \rm av \infty}=\frac{1}{4 \pi}
\int_{4 \pi} \dd \Omega_{\bm{k}} \, H_E^{\infty},
\label{e:Hav}
\end{equation}
where $\dd \Omega_{\bm{k}}$ is a solid angle element of possible orientations $\bm{k}$
of the detector on the sky. The angle-averaging simplifies the integration,
\begin{equation}
H_E^{\rm av \infty}=\frac{15 \SB }{16 \pi^5 {k_{\rm B}^4}}\, 
\int_{4 \pi} \dd \Omega_{\rm s}\,
\frac{(1-\xg)^{-1}\,  E^3}{\exp(E/\kB \Ts^\infty)-1};
\label{e:Hav1} 
\end{equation} 
the angle-averaged bolometric flux determines the total
luminosity~(\ref{e:Ls}) of the star, 
$L_{\rm s}^\infty=4 \pi D^2 F_{\rm bol}^{\rm av\infty}= 4 \pi \SB R_\infty^2 \Teff^{\infty 4}$.

For a uniform surface temperature distribution, 
one naturally recovers the ordinary BB($\equiv$1BB) law,
\begin{equation}
H_E^{\rm BB\infty}=\frac{15 \SB }{4 \pi^4 k_{\rm B}^4}\,
\frac{(1-\xg)^{-1}\, E^3}{\exp(E/\kB \Teff^\infty)-1},
\label{e:BB} 
\end{equation} 
and the bolometric effective flux 
\begin{equation}
H_{\rm bol}^{\rm BB\infty}=\int_0^\infty H_E^{\rm BB \infty}\,
\dd E=\frac{\SB \Teff^{\infty 4}}{1-\xg},
\label{e:BBbol}
\end{equation}
with  
$L_{\rm s}^\infty=4 \pi D^2 F_{\rm bol}^{\rm BB\infty}= 4 \pi \SB R_\infty^2 \Teff^{\infty 4}$.

\subsection{Illustrative examples}
\label{s:Palex model}

\begin{figure}
	\includegraphics[width=0.45\textwidth]{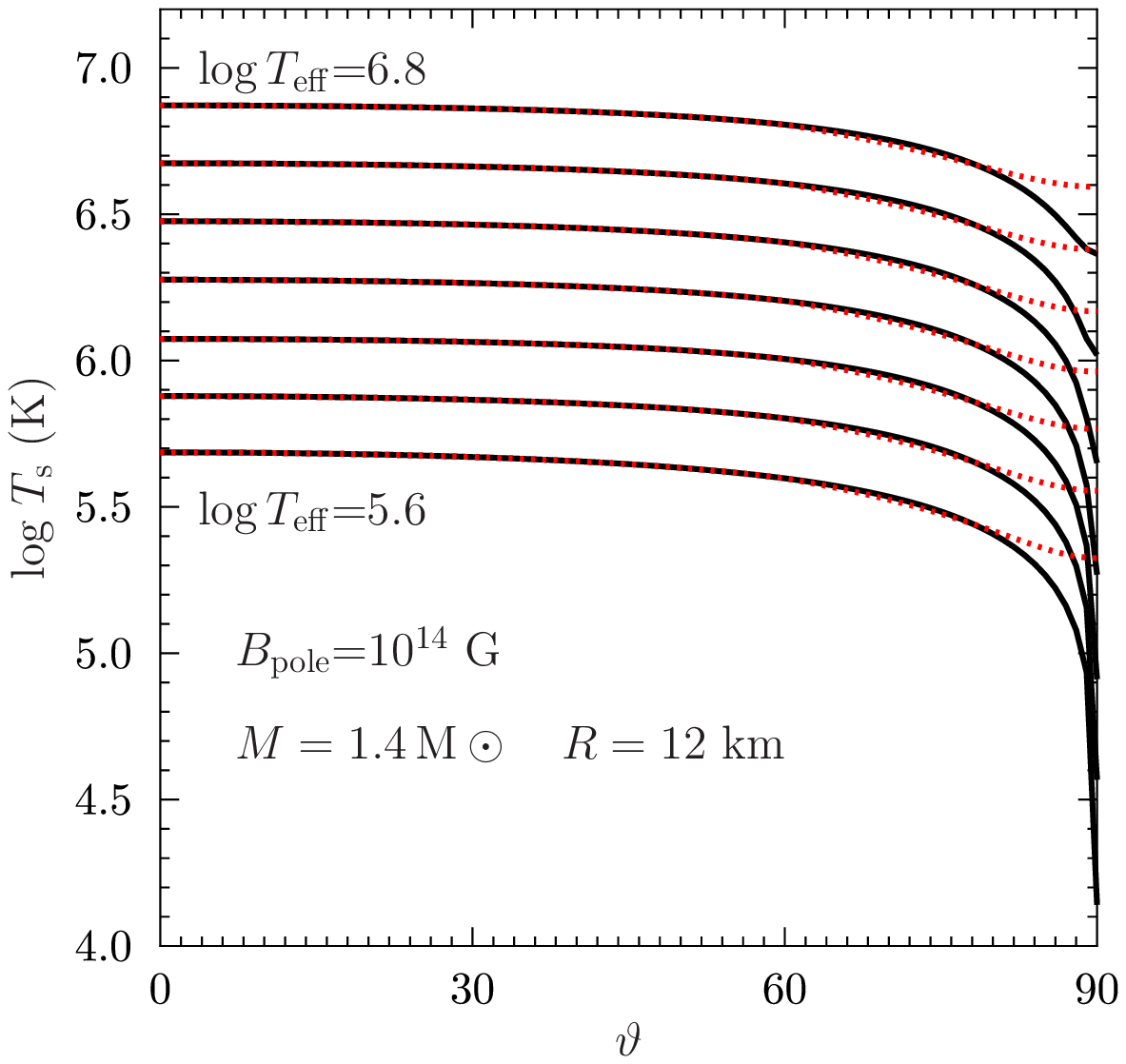}%
	\caption{
		Logarithm of the surface temperature $\Ts$ versus colatitude
		$\vartheta$, from magnetic pole ($\vartheta=0$) to equator 
		($\vartheta=90^\circ$) of a neutron star with 
		$M=1.4\,\msun$ and $R=12$ km at $\Bpole=10^{14}$~G. 
		The curves are plotted for seven effective temperatures, 
		$\log \Teff$ [K]=5.6, 5.8, \ldots 6.8. Solid curves 
		refer to the basic heat-blanket model \citep{PY01,PYCG03,PC18},
		while dots display an artificial model discussed in Sect.\ \ref{s:coldbelt}.
		See the text for details.		
	}
	\label{f:Rsoftheta}
\end{figure}

For illustration, Fig.~\ref{f:Rsoftheta} shows (by solid lines) 
the surface temperature distribution
for a passively cooling magnetized neutron star with pure dipole magnetic field
in the outer layers, assuming $M=1.4\, \msun$ and
$R=12$ km [$\xg$=0.344 and the surface gravity
$g_{\rm s}=GM/(R^2 \sqrt{1-\xg})=1.59 \times 10^{14}$ cm~s$^{-2}$]. 
The surface field at the magnetic pole is $\Bpole=10^{14}$~G.
The plots are 
given by the standard theory 
of magnetized heat blanketing envelopes (e.g., \citealt{PY01,PYCG03,PC18,BPY21}).
The $\Ts$ distributions are axially symmetric with respect to the magnetic axis and 
equator. It is sufficient to plot $\Ts$ versus colatitude $\vartheta$ 
from $\vartheta=0$ (magnetic pole)
to $90^\circ$ (equator). We present the solid curves for seven effective
temperatures $\Teff$. The respective dotted curves are for a toy model
of heat blankets considered in  Sect.\ \ref{s:coldbelt}.

The star is assumed to be isothermal inside. The main temperature gradient
and the $\Ts (\bm{n})$ distribution (solid lines) are mediated by
heat transport through a thin outer heat blanketing envelope, which extends
from the surface to the density $\rhob \sim 10^{10}$ \gcc. 
For certainty, unless the contrary is indicated, the blanketing envelope
is assumed to be made of iron. The magnetic field
makes heat transport anisotropic and creates variations of $\Ts(\bm{n})$, which become noticeable at $\Bpole \gtrsim 10^{11}$ G.
In Fig.~\ref{f:Rsoftheta} displays strong pole-to-equator $\Ts$ variations (solid lines)
exceeding one order of magnitude at $\log \Teff$ [K]=5.6. The pole is hotter than
the equator because of higher heat conduction in the polar regions, where
the heat outflows to the surface nearly along magnetic field lines. In contrast,
there is a colder equatorial belt, where the heat outflows nearly across magnetic
lines; such outflow is strongly suppressed. 
In a hotter star with higher $\Teff$, the $\Ts$ anisotropy
is less pronounced due to reduction of heat transport anisotropy in
warmer matter. With increasing $\Bpole$, the polar regions become warmer, while
the equatorial belt colder and thinner.

\begin{figure}
	\includegraphics[width=0.45\textwidth]{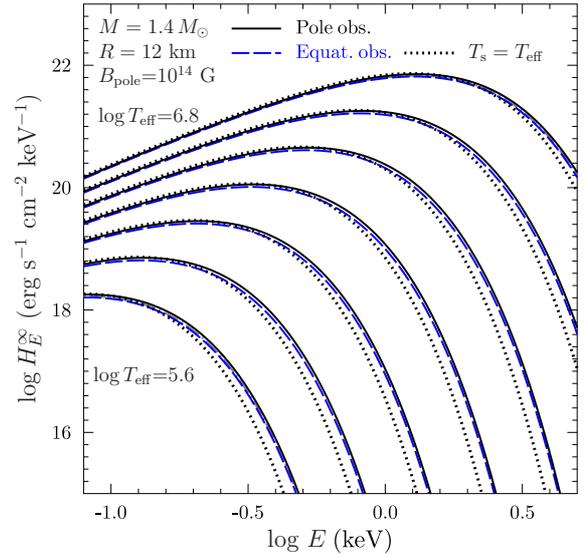}%
	\caption{
		Effective spectral flux density $H^\infty_E$ versus
		redshifted photon energy $E$ at (from bottom to top) 
		$\log \Teff$ [K]=5.6, 5.8, \ldots 6.8 from
		a $1.4\,\msun$ neutron star with $R=12$ km and $\Bpole=10^{14}$~G (as in Fig.\ \ref{f:Rsoftheta}). 
		Solid lines refer to pole observations (inclination angle $i=0$), 
		dashed lines to equator observations ($i=90^\circ$);
		dotted lines are for the 
		1BB model with given $\Teff$.		
	}
	\label{f:basic}
\end{figure}

\begin{figure}
	\includegraphics[width=0.45\textwidth]{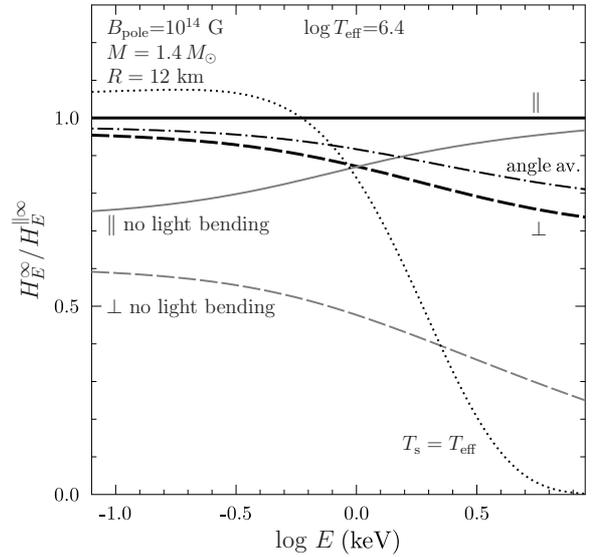}%
	\caption{
		Effective spectral flux density $H^\infty_E$ 
		in units of $H^{\parallel \infty}_E$ versus
		redshifted photon energy $E$ at  
		$\log \Teff$ [K]=6.4 for
		a $1.4\,\msun$ neutron star with $R=12$ km and $\Bpole=10^{14}$.  
		Solid lines refer to pole observations, dashed lines to equator observations;
		dot-dashed line shows the angle-averaged flux, while the dotted line is for the 
		1BB ($\Ts=\Teff$) model. Grey solid and dashed lines
		are the same fluxes $H^{\parallel \infty}_E$ and $H^{\perp \infty}_E$, as the black
		lines, but neglecting the gravitational light bending of photons. 		
	}
	\label{f:compar}
\end{figure}

Fig. \ref{f:basic} 
shows the thermal spectral flux density $H_E^\infty$ 
emitted by same star as in Fig.\ \ref{f:Rsoftheta} (for the same seven
values of $\Teff$ diplayed by solid lines). The flux is plotted
in logarithmic scale as a function of decimal logarithm of the redshifted photon energy $E$.  For each value of $\Teff$ we present three curves.  
The solid curves give the flux
$H_E^{\parallel \infty}$, as
measured in pole observations (marked as $\parallel$,
with the inclination $i=0$ to the magnetic axis).
The dashed curves present the flux $H_E^{\perp \infty}$ for
the equator observations (marked with $\perp$, at
$i=90^\circ$). 
The dotted curves demonstrate 
the plain 1BB model, equation (\ref{e:BB}), with given $\Ts=\Teff$ 
constant over the surface.
The angle-averaged curves $H_E^{\rm av \infty}$, given by
equation (\ref{e:Hav}), always lie between
$H_E^{\parallel \infty}$ and $H_E^{\perp \infty}$ and would be 
indistinguishable from the dotted curves in the logarithmic format. 

All four fluxes for each $\Teff$ look
very close in the logarithmic format of Fig.\ \ref{f:basic}.
To demonstrate their difference, 
Fig.\ \ref{f:compar} plots spectral fluxes $H^\infty_E$ in natural 
 format. Here we
take $\log \Teff=6.4$ and express all  $H^\infty_E$ in units of
$H^{\parallel \infty}_E$. This normalization gives  
the unit relative flux (solid black line)
for pole observations. The flux for the equator observations (long--dashed black line)
is always lower because the observer detects radiation emitted from overall colder
regions. The $H^{\perp \infty}_E/H^{\parallel \infty}_E$ ratio is seen to decrease with photon
energy. The black dash--dot line refers to $H^{\rm av \infty}_E$, while the
dotted black line to $H^{\rm BB \infty}_E$. One can see that these lines become noticeably different
at higher $E$.  Finally, the grey solid and long--dash
curves refer to the same $H^{\parallel \infty}_E$ and $H^{\perp \infty}_E$ fluxes, as
the black curves, but calculated neglecting gravitational light bending. These grey curves
are systematically lower than their black counterparts because the neglect of light bending
decreases observable surface area of the star. This confirms the importance of
light bending.
Note also that the light bending greatly reduces the difference of fluxes in the pole and
equator observations, making the thermal neutron star radiation more isotropic \citep{page95}.

\section{2BB approximation}
\label{s:2BB}

\subsection{2BB fits}
\label{s:2BBfits}

Here we show that the multi--BB fluxes discussed in Sect.\
\ref{s:Palex model} can be accurately fitted by a simple 2BB model. The idea is not new (e.g. \citealt{popov17}) 
but, to the best of our knowledge, it has not been studied
in great details. Let us 
take the same model of a spherical neutron star with a dipole magnetic
field near the surface.  
We have calculated the spectral flux $H_E^\infty$ from equation (\ref{e:H(Einfty)})
for a number of values of $M$, $R$, $\Bpole$, $\Teff$ and inclination angles
$i$. We have mainly 
considered photon  energies $E$ from 0.064 keV
to $E_{\rm max}=42.49$ keV 
(overlapping thus typical observable X-ray range)
on a grid of 142 energy points placed logarithmically equidistant within
this range.  The spectra have been fitted by the 2BB model,
\begin{equation}
	H_E^\infty=\scc H_E^{\rm BB \infty}(T_{\rm eff c})+
	\shh H_E^{\rm BB \infty}(T_{\rm eff h}).
	\label{e:2BBfit}    
\end{equation}
Here $H_E^{\rm BB \infty}(\Teff)$ is given by equation 
(\ref{e:BB}), with $T_{\rm effc}=\pcc \Teff$ and $T_{\rm effh}=\phh \Teff$; the
subscripts `c' and `h' refer to the colder and the hotter BB components, respectively. 
Four positive dimensionless fit parameters $\scc$, $\shh$, $\pcc$ and $\phh$ 
define relative contributions of the two 
BB components to the total
thermal flux $H_E^\infty$. The parameters $\pcc$ and $\phh$ determine
the effective temperatures of the components, while
$\scc$ and $\shh$ are effective fractions of their surface areas
(without imposing $\scc+\shh=1$). 
Since the first component is colder, we have $\pcc<\phh$. We treat $s_\text{c,h}$ and 
$p_\text{c,h}$ just as formal fit parameters (not the parameters of real
colder and hotter uniform
areas on the neutron star surface). Very roughly, one can visualize $R_{\rm c}=R\,\sqrt{s_{\rm c}}$ 
and $R_{\rm h}=R\,\sqrt{s_{\rm h}}$ as the  effective radii of the colder and hotter
emission regions, respectively.
All four fit parameters $\scc$, $\shh$, $\pcc$ and $\phh$ 
depend on $M$, $R$, $\Bpole$, $\Teff$ and $i$. Actually, these fit parameters are 
not independent. They should guarantee the required bolometric luminosity
of the star, that is determined by given  $\Teff$ in accord with equation (\ref{e:Ls}).
We used this equation to check the quality of our 2BB fits and
it always worked well.

\subsection{2BB fits: pole and equator observations}
\label{s:2BBfit}

We start with the two basic models for pole and
equator observations ($H_E^{\parallel \infty}$ and $H_E^{\perp \infty}$,
$i=0$ and 90$^\circ$), leaving the case of arbitrary $i$ to
Sect.\ \ref{s:2BBfit-anyi}. 

For example, Table \ref{tab:fit}
presents the fit results for the star with $M=1.4$ $\msun$ and $R=12$ km.
The fitting has been performed on the indicated grid of 142 energy points $E$.
Before fitting, we have removed the energy points,  
at which $H_E^\infty<1$ erg s$^{-1}$ cm$^{-2}$ keV$^{-1}$ and the spectral flux is negligible.
Table \ref{tab:fit} gives the best fit parameters $\pcc$, $\phh$, $\scc$ and $\shh$
for magnetic pole ($\parallel$) and equator ($\perp$) observations at
four values of $\Bpole=10^{11}$, $10^{12}$, $10^{13}$ and $10^{14}$~G.
For each $\Bpole$, we consider $6 \times 2$ atmosphere
models with $\log \Teff$ [K]=5.6, 5.8,\ldots,6.6 (6 lines with eight fit 
parameters for any $\Teff$ in Table \ref{tab:fit}). In each line, we present also
decimal logarithm of the internal temperature $\log \Tb$ and two root mean squared
relative fit errors $\delta_{\rm rms}$ (for the two observation directions,  $\parallel$ and
$\perp$). One can see that in all 48 fit cases $\delta_{\rm rms}$ does not exceed
2.5 per cent. The maximum relative fit error is typically  $\approx (2-2.5)\, \delta_{\rm rms}$. 
Such fits can be regarded as nearly perfect (`too accurate'), taking into account 
simplified nature of the initially calculated  fluxes. If plotted on Figs.~\ref{f:basic} and \ref{f:compar},
these fits would be indistinguishable from corresponding solid and long--dashed curves.

\renewcommand{\arraystretch}{1.1}
\begin{table*}
	\caption{Fit parameters $\pcc$, $\phh$, $\scc$ and $\shh$ in equation (\ref{e:2BBfit}), together with
		rms relative fit errors $\delta_{\rm rms}$,  
		for a neutron star with $M=1.4$ $\msun$, $R=12$ km at two observation directions 
		($\parallel$ and $\perp$), four values of $\log \Bpole$ [G]=11, 12, 13, 14, and
		six values of $\log \Teff$ [K]=5.6, 5.8, \ldots, 6.6 (with corresponding
		logarithms of the internal temperature,  $\log \Tb$ [K]). 
	}
	\label{tab:fit}
	\begin{tabular}{c c c c c c c c c c c c c}
		\hline 
		$\log \Bpole$ &  $\log \Ts$   & $\log \Tb$  & $p_{\rm c\parallel}$  & $p_{\rm h\parallel}$  & $s_{\rm c\parallel}$ & $s_{\rm h\parallel}$ & $\delta_{\rm rms}^\parallel$ & $p_{\rm c\perp}$  & $p_{\rm h \perp}$  & $s_{\rm c\perp}$ & $s_{\rm h\perp}$ & $\delta_{\rm rms}^\perp$  \\
		\hline
		11 & 5.6 &  7.3854 &  0.9708 &  1.1262 &  0.5126 &  0.3669 &  0.011 &  0.9624 &  1.1239 &  0.5620 &  0.3002 &  0.014 \\
		11 & 5.8 &  7.7477 &  0.9552 &  1.1014 &  0.5641 &  0.3898 &  0.009 &  0.9493 &  1.0991 &  0.6258 &  0.3218 &  0.011 \\
		11 & 6.0 &  8.0979 &  0.9591 &  1.0750 &  0.5793 &  0.4062 &  0.005 &  0.9553 &  1.0729 &  0.6453 &  0.3384 &  0.006 \\
		11 & 6.2 &  8.4447 &  0.9685 &  1.0506 &  0.5696 &  0.4269 &  0.002 &  0.9663 &  1.0488 &  0.6360 &  0.3601 &  0.002 \\
		11 & 6.4 &  8.7918 &  0.9775 &  1.0313 &  0.5426 &  0.4569 &  0.001 &  0.9761 &  1.0297 &  0.6046 &  0.3950 &  0.001 \\
		11 & 6.6 &  9.1410 &  0.9860 &  1.0191 &  0.5309 &  0.4690 &  0.000 &  0.9853 &  1.0180 &  0.5931 &  0.4069 &  0.000 \\
		\hline
		 12 & 5.6 &  7.3488 &  1.0057 &  1.1628 &  0.4465 &  0.3312 &  0.011 &  0.9952 &  1.1605 &  0.4757 &  0.2667 &  0.014 \\
		 12 & 5.8 &  7.7410 &  0.9845 &  1.1572 &  0.4682 &  0.3526 &  0.014 &  0.9743 &  1.1546 &  0.5055 &  0.2863 &  0.018 \\
		 12 & 6.0 &  8.1185 &  0.9622 &  1.1490 &  0.4945 &  0.3734 &  0.017 &  0.9523 &  1.1463 &  0.5406 &  0.3060 &  0.021 \\
		 12 & 6.2 &  8.4868 &  0.9435 &  1.1359 &  0.5281 &  0.3913 &  0.017 &  0.9348 &  1.1331 &  0.5836 &  0.3231 &  0.021 \\
		 12 & 6.4 &  8.8435 &  0.9398 &  1.1158 &  0.5582 &  0.4022 &  0.013 &  0.9330 &  1.1132 &  0.6209 &  0.3335 &  0.016 \\
		 12 & 6.6 &  9.1898 &  0.9485 &  1.0895 &  0.5735 &  0.4108 &  0.007 &  0.9440 &  1.0872 &  0.6397 &  0.3423 &  0.009 \\
		\hline
		13 & 5.6 &  7.2925 &  1.0032 &  1.1590 &  0.4312 &  0.3461 &  0.011 &  0.9933 &  1.1567 &  0.4616 &  0.2799 &  0.013 \\
		13 & 5.8 &  7.6673 &  0.9901 &  1.1646 &  0.4613 &  0.3434 &  0.014 &  0.9794 &  1.1621 &  0.4955 &  0.2776 &  0.018 \\
		13 & 6.0 &  8.0568 &  0.9767 &  1.1642 &  0.4818 &  0.3500 &  0.017 &  0.9658 &  1.1616 &  0.5203 &  0.2840 &  0.021 \\
		13 & 6.2 &  8.4502 &  0.9621 &  1.1590 &  0.4923 &  0.3659 &  0.019 &  0.9515 &  1.1563 &  0.5360 &  0.2989 &  0.023 \\
		13 & 6.4 &  8.8348 &  0.9476 &  1.1527 &  0.5025 &  0.3824 &  0.020 &  0.9375 &  1.1499 &  0.5513 &  0.3146 &  0.024 \\
		13 & 6.6 &  9.2082 &  0.9347 &  1.1452 &  0.5176 &  0.3968 &  0.019 &  0.9253 &  1.1422 &  0.5719 &  0.3284 &  0.024 \\
		\hline
		14 & 5.6 &  7.1730 &  1.0225 &  1.1914 &  0.4613 &  0.2819 &  0.013 &  1.0095 &  1.1891 &  0.4807 &  0.2230 &  0.016 \\
		14 & 5.8 &  7.6315 &  0.9956 &  1.1718 &  0.4702 &  0.3268 &  0.015 &  0.9840 &  1.1692 &  0.5017 &  0.2629 &  0.018 \\
		14 & 6.0 &  7.9913 &  0.9742 &  1.1603 &  0.4688 &  0.3629 &  0.016 &  0.9637 &  1.1576 &  0.5084 &  0.2958 &  0.020 \\
		14 & 6.2 &  8.3648 &  0.9650 &  1.1644 &  0.4896 &  0.3589 &  0.019 &  0.9539 &  1.1616 &  0.5315 &  0.2922 &  0.023 \\
		14 & 6.4 &  8.7685 &  0.9561 &  1.1632 &  0.5022 &  0.3636 &  0.020 &  0.9449 &  1.1603 &  0.5469 &  0.2968 &  0.025 \\
		14 & 6.6 &  9.1668 &  0.9456 &  1.1580 &  0.5064 &  0.3774 &  0.020 &  0.9350 &  1.1550 &  0.5547 &  0.3099 &  0.025 \\		
		\hline
	\end{tabular}
	\\
\end{table*}
\renewcommand{\arraystretch}{1.0} 

\subsection{2BB fits: any observation angle}
\label{s:2BBfit-anyi}

In case of arbitrary
inclination $i$, 
the spectral flux $H_E^\infty=H_E^{i\infty}$ can be computed from equation (\ref{e:H(Einfty)}).
We have performed such computations for a range of values of $M$, $R$, $\Bpole$, $\Teff$ and 
$i$. In all the cases the computed fluxes $H^{i \infty}_E$
are nicely approximated by
\begin{equation}
	H^{i \infty}_E=H_E^{\parallel \infty} \cos^2 i + H_E^{\perp \infty} \sin^2 i,
	\label{e:Hanyi}
\end{equation}
where $H_E^{\parallel \infty}$ and $H_E^{\perp \infty}$ are the fluxes for the
pole and equator observations, discussed above. The fit accuracy of equation (\ref{e:Hanyi}) is of the 
same order of magnitude as the 2BB fit accuracy of equation (\ref{e:2BBfit}).
Therefore, any thermal spectral flux $H^{i \infty}_E$ observed at any direction $i$ is determined by
the two basic pole and equatorial spectral fluxes, $H_E^{\parallel \infty}$ and
$H_E^{\perp \infty}$, which are accurately described by 2BB fits with easily
computable fit parameters (Table \ref{tab:fit}). This gives practical solution for
the assumed model of magnetized neutron star emission. Note that equation
(\ref{e:Hanyi}) is the same as the equation for the flux produced by a classic electric dipole emission in flat space.

\subsection{2BB phase--space maps}   
\label{s:maps}

\begin{figure*}
	\includegraphics[width=0.45\textwidth]{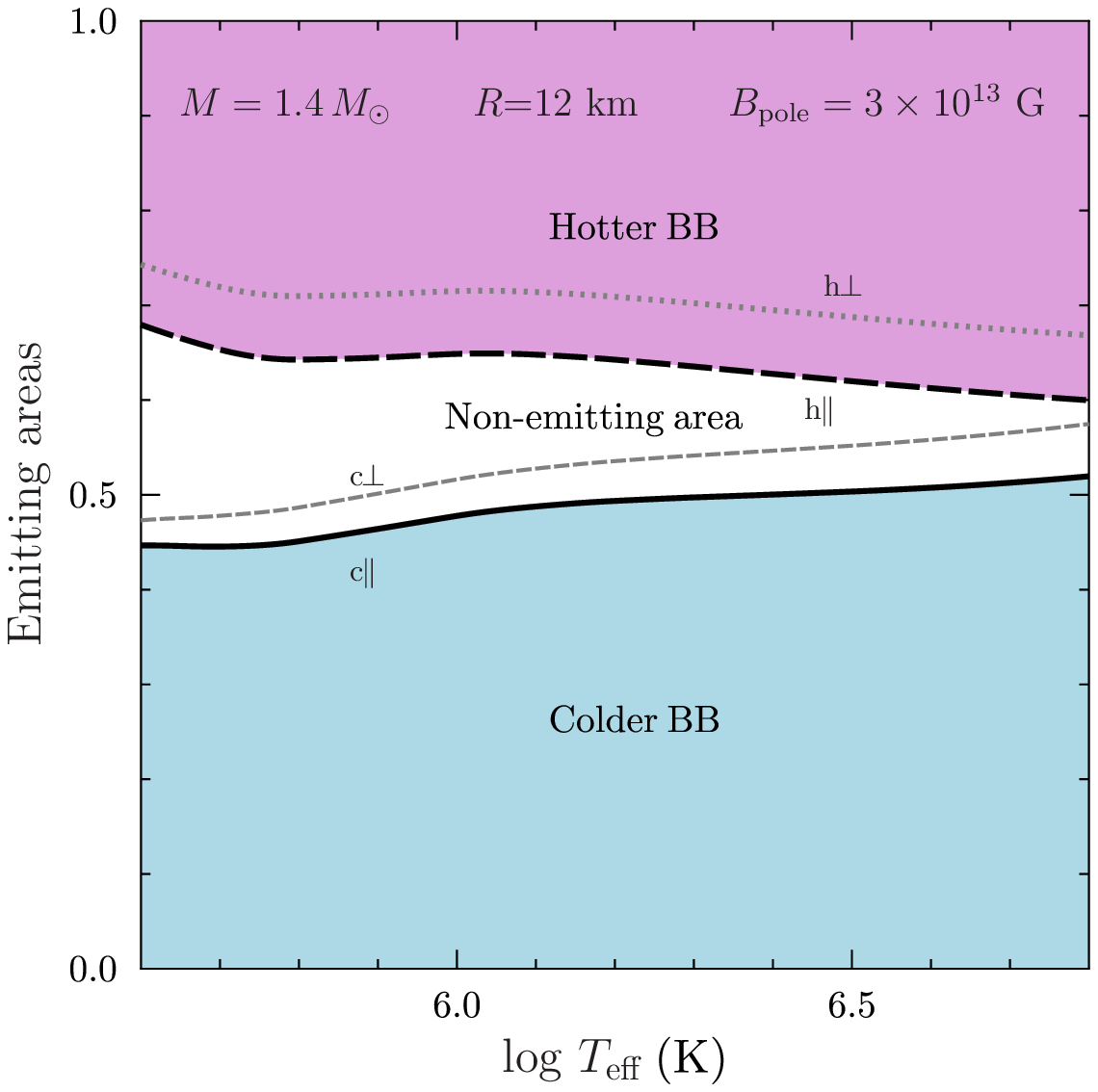}%
	\hspace{5mm}
	\includegraphics[width=0.45\textwidth]{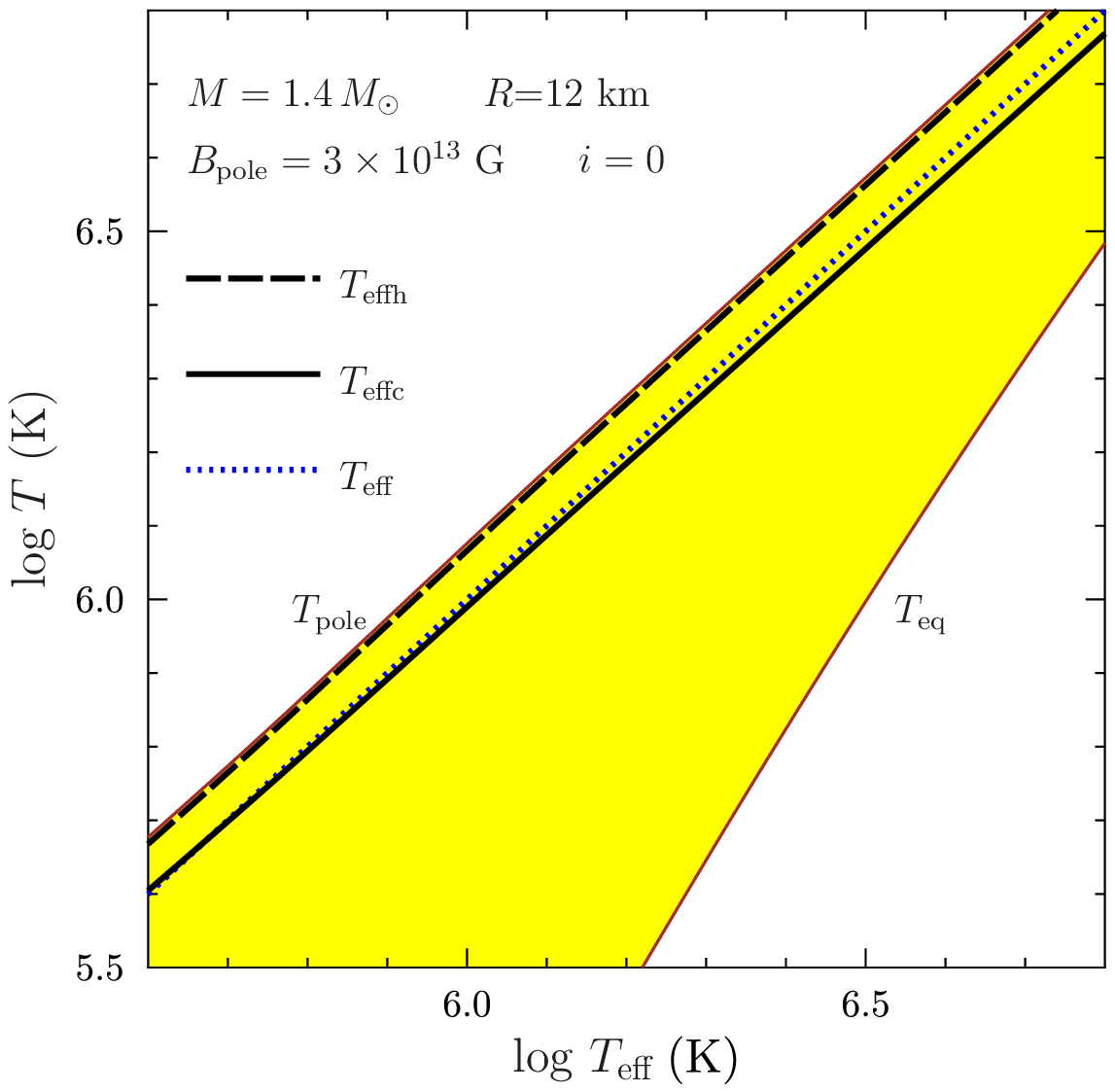}%
	\caption{
		Map of 2BB fit parameters versus  
		$\log \Teff$ for
		a $1.4\,\msun$ neutron star with $R=12$ km and $\Bpole=3 \times 10^{13}$~G for
		pole observation ($i=0$). The left--hand panel shows the fractional surface areas
		$s_{\rm h}$ and $s_{\rm c}$ for the hotter and colder BB components as lengths
		of segments of vertical lines at given $\Teff$ in the upper and lower shaded zones,
		respectively. The blank intermediate zone  gives `non-emitting' fractional
		surface area. The short--dashed and dotted grey lines show the
		boundaries of the shaded zones for the equator observation ($i=90^\circ$). The right--hand
		panel demonstrates characteristic temperatures. The shaded zone shows range of
		$\Ts$--variations over the surface, from the highest temperature $T_{\rm pole}$
		at the pole to the lowest one $T_{\rm eq}$ at the equator. The dotted line is 
		the total effective temperature $\Teff$, while the dashed and solid lines
		are the effective temperatures $T_{\rm effh}$ 
		and $T_{\rm effc}$ for the hotter and colder BB components (which are almost insensitive
		of $i$).
		See the text for details.	
	}
	\label{f:maps}
\end{figure*}

\begin{figure*}
	\includegraphics[width=0.42\textwidth]{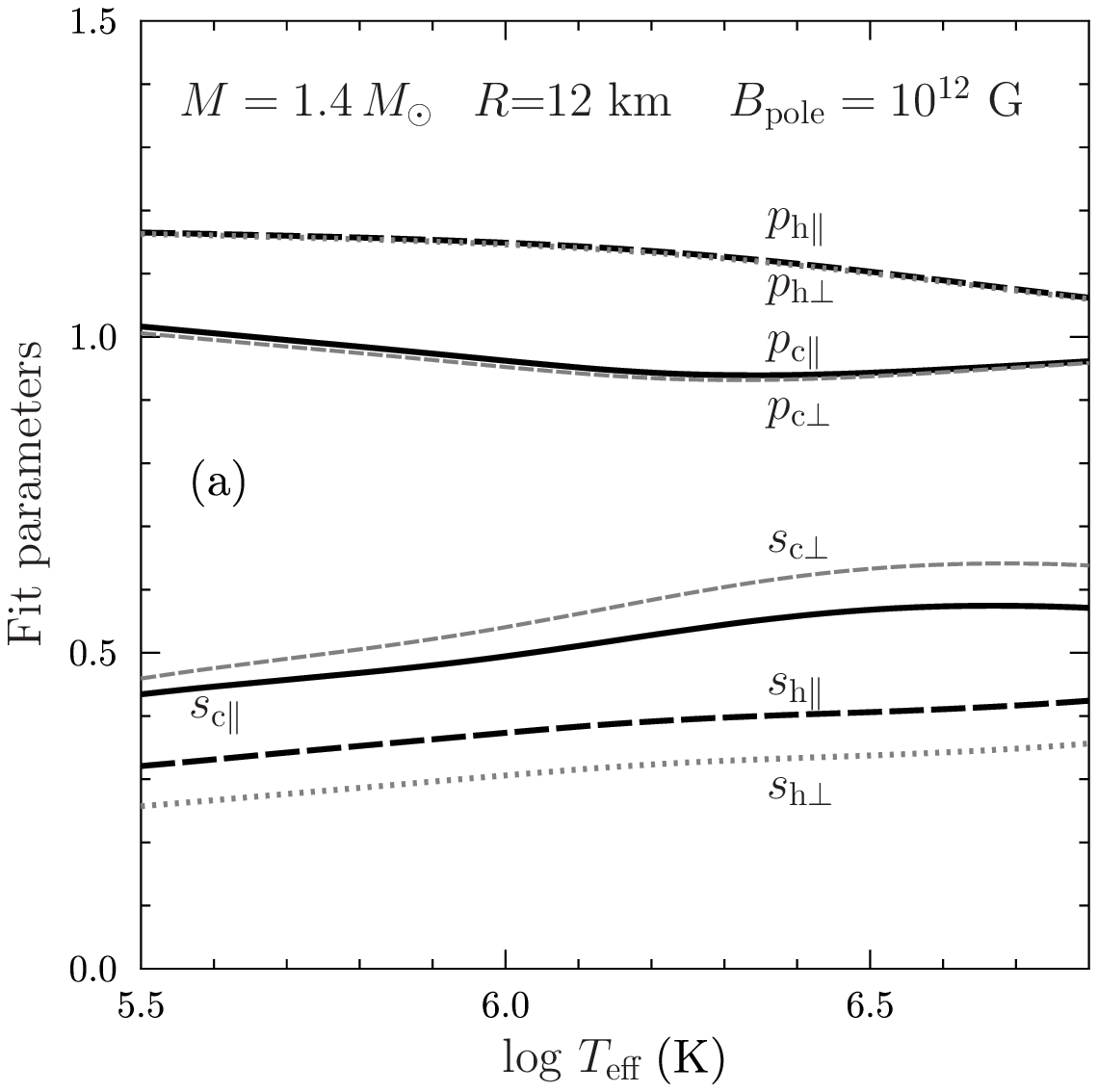}%
	\hspace{1mm}
	\includegraphics[width=0.42\textwidth]{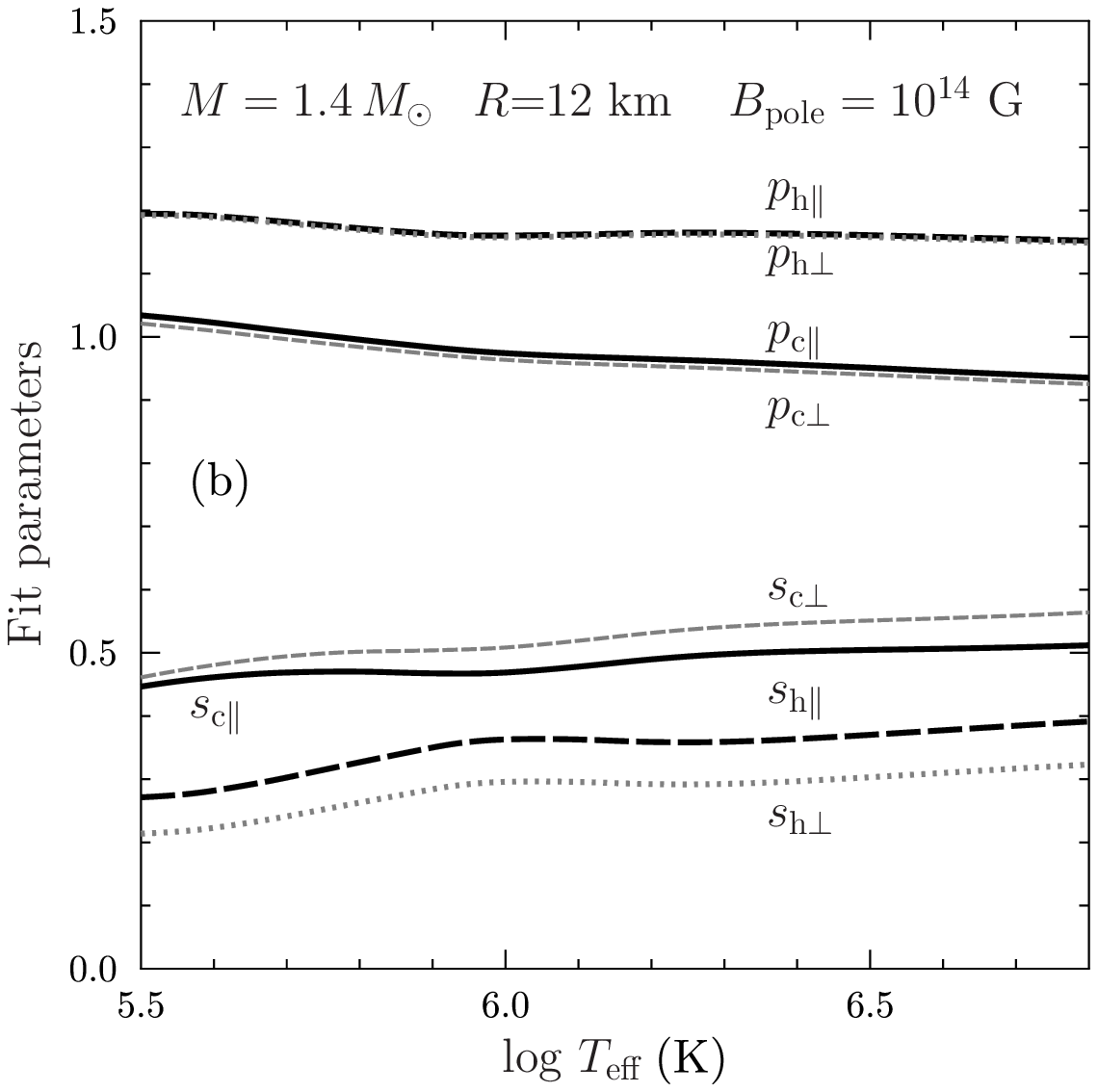}%
	\caption{
		2BB fit parameters versus  
		$\log \Teff$ for
		a $1.4\,\msun$ neutron star with $R=12$ km and $\Bpole=10^{12}$ [panel (a)]
		or $10^{14}$~G [panel (b)]. 
		Black and grey lines refer, respectively, to pole and
		equator observations.
		Black solid lines show $p_{\rm c \parallel}$ and $s_{\rm c \parallel}$ (for the colder BB components),
		while black long--dashed lines show $p_{\rm h \parallel}$ and $s_{\rm h \parallel}$ 
		(for the hotter BB components). Short--dashed grey lines refer to $p_{\rm c \perp}$ and $s_{\rm c \perp}$, 
		whereas dotted grey lines to  $p_{\rm h \perp}$ and $s_{\rm h \perp}$. 	
	}
	\label{f:maps1412}
\end{figure*}

\begin{figure*}
	\includegraphics[width=0.42\textwidth]{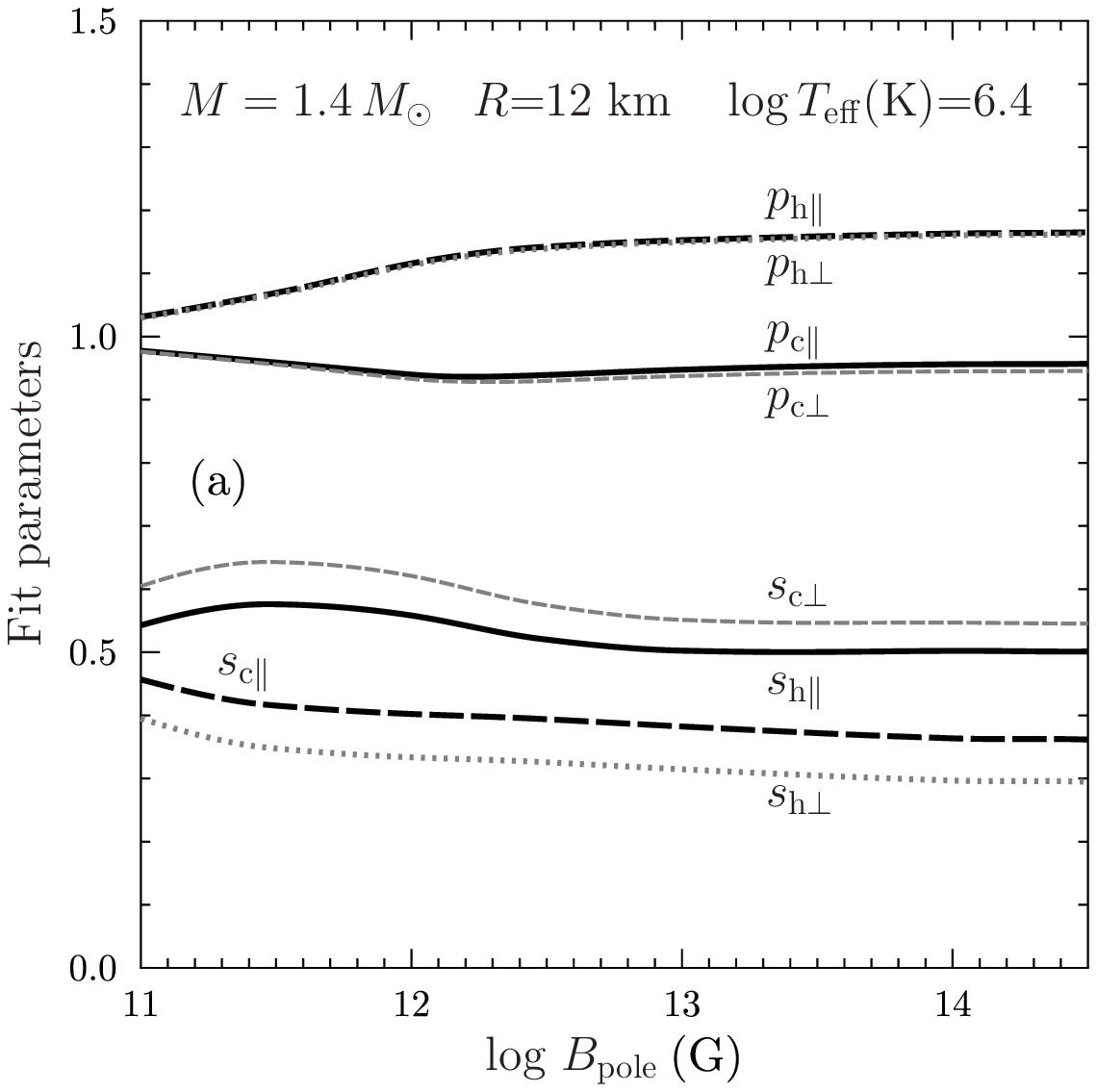}%
	\hspace{1mm}
	\includegraphics[width=0.42\textwidth]{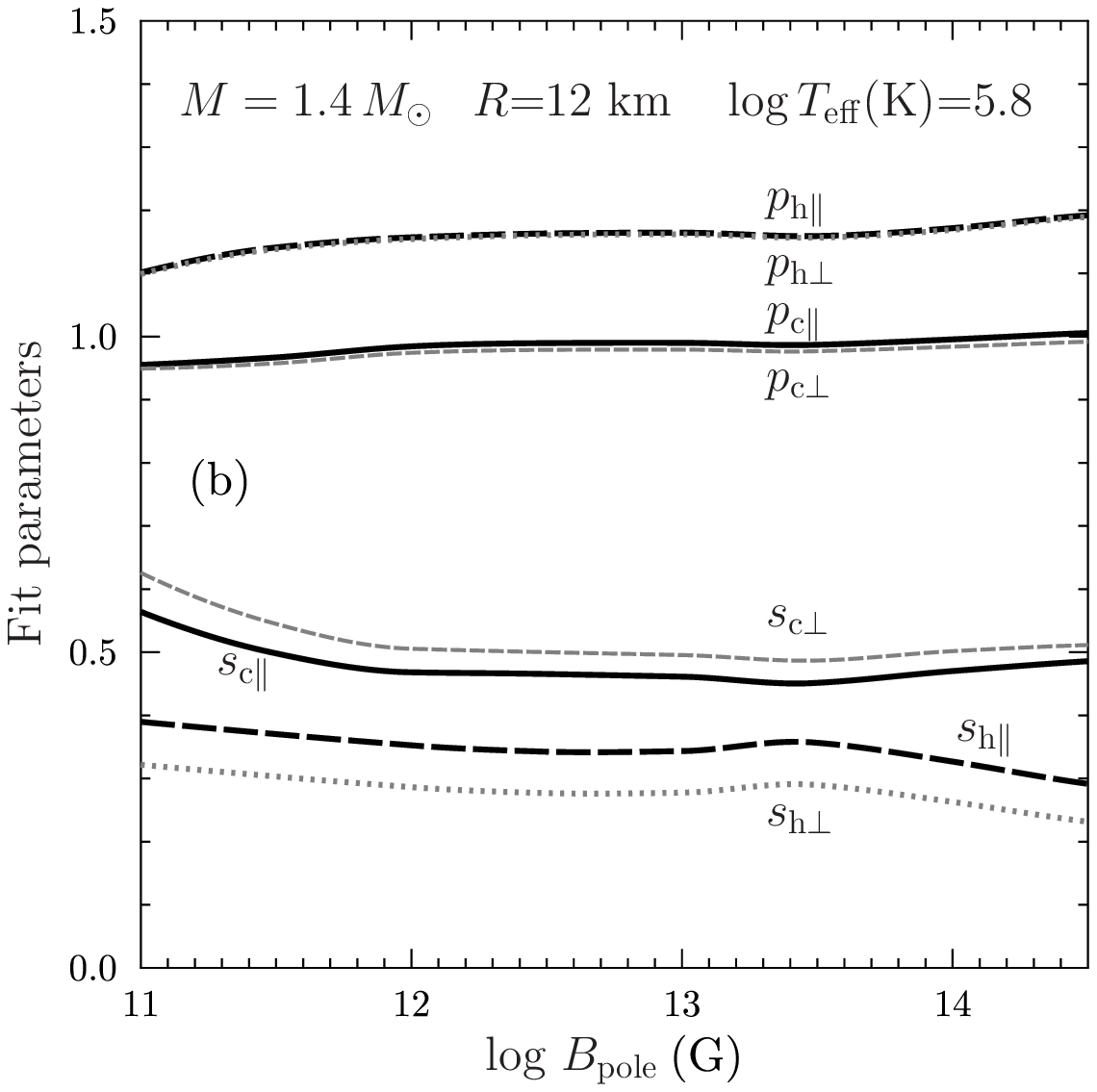}%
	\caption{
		Same 2BB fit parameters as in Fig.\ \ref{f:maps1412} 
		for
		a $1.4\,\msun$ neutron star with $R=12$ km 
		but versus $\Bpole$ at
		$\log \Teff$[K]=6.4 [panel (a)]
		and $5.8$ [panel (b)].     
		See the text for details.	
	}
	\label{f:mapsofB}
\end{figure*}

Many observed thermal X-ray spectra of neutron stars have been fitted by 2BB models.
Since 2BB models accurately approximate a wide class of multi--BB models, it seems 
attractive to develop a technique for studying properties of magnetized stars,
as multi--BB emitters, using already available 2BB fits of the observed spectra. This can be done by comparing theoretical
2BB maps (2BB parameters calculated in a wide phase--space of input parameters,
such as $M$, $R$, $\Bpole$, $\Teff$, $i$), with observations and selecting then most
suitable phase--space domains.

Some maps are demonstrated in Fig.\ \ref{f:maps}. The figure refers
to a $1.4 \msun$ star with $R=12$ km and $\Bpole=3 \times 10^{13}$ G, observed from
the pole. The fit parameters are displayed versus $\Teff$. The left--hand side plots emitting
fractions of the cold and hot BB surface areas, $s_{\rm c}$ and  $s_{\rm h}$
(lower and upper shaded areas, respectively). For any $\Teff$ we have 
$s_{\rm c}+s_{\rm h}<1$, which may be viewed as the presence 
of `non-emitting' fraction of the stellar surface shown as blank. This effect supports
common knowledge that one should be careful in using  $s_{\rm c}$ 
and  $s_{\rm h}$ for estimatimg true neutron star radius from observations.
Notice that the presence of the `dark' surface does not affect
excellent quality of the 2BB fits. 
With increasing $\Teff$, the fraction of BB 'dark' surface decreases.

The right--hand panel of Fig.~\ref{f:maps} shows characteristic temperatures  
of the emitting star. The shaded zone is restricted by the maximum temperature 
$T_{\rm pole}$ at the magnetic pole and the minimum temperature $T_{\rm eq}$ 
at the equator. We see that temperature variations over the surface of
a magnetized star are large. The dotted curve plots the mean effective
temperature $\Teff$ (to guide the eye). The dashed and solid courves are the effective
BB temperatures, $T_{\rm effh}$ and $T_{\rm effc}$, respectively.

Although Fig.~\ref{f:maps} is designed for the pole observations, the grey lines
on the left-hand panel show the deformations of the emitting zones for the 
equator observations; these deformations are not dramatic but noticeable.
Corresponding deformations on the right--hand panel are not plotted; they
would be almost invisible.

\begin{figure*}
	\includegraphics[width=0.42\textwidth]{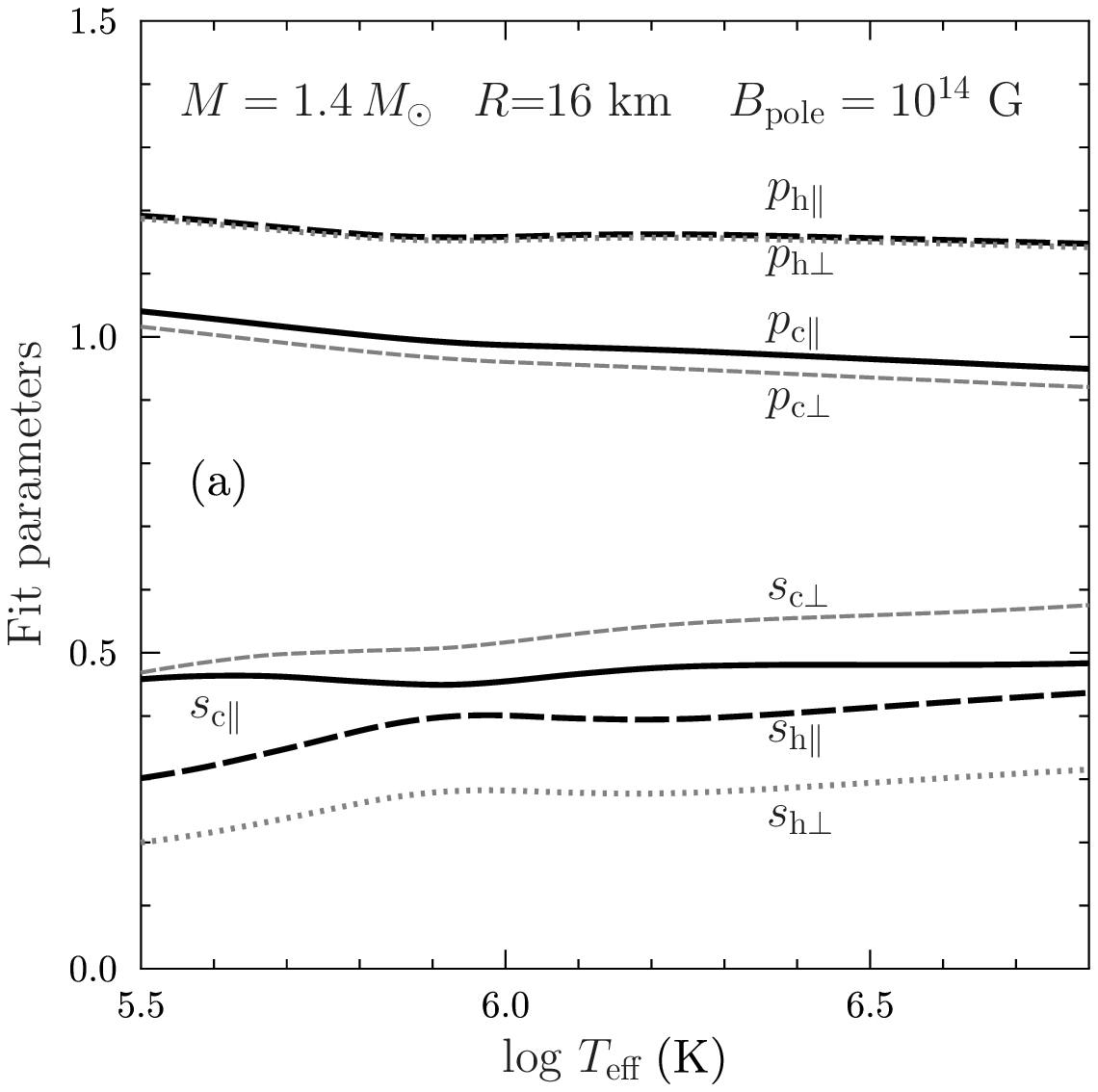}%
	\hspace{1mm}
	\includegraphics[width=0.42\textwidth]{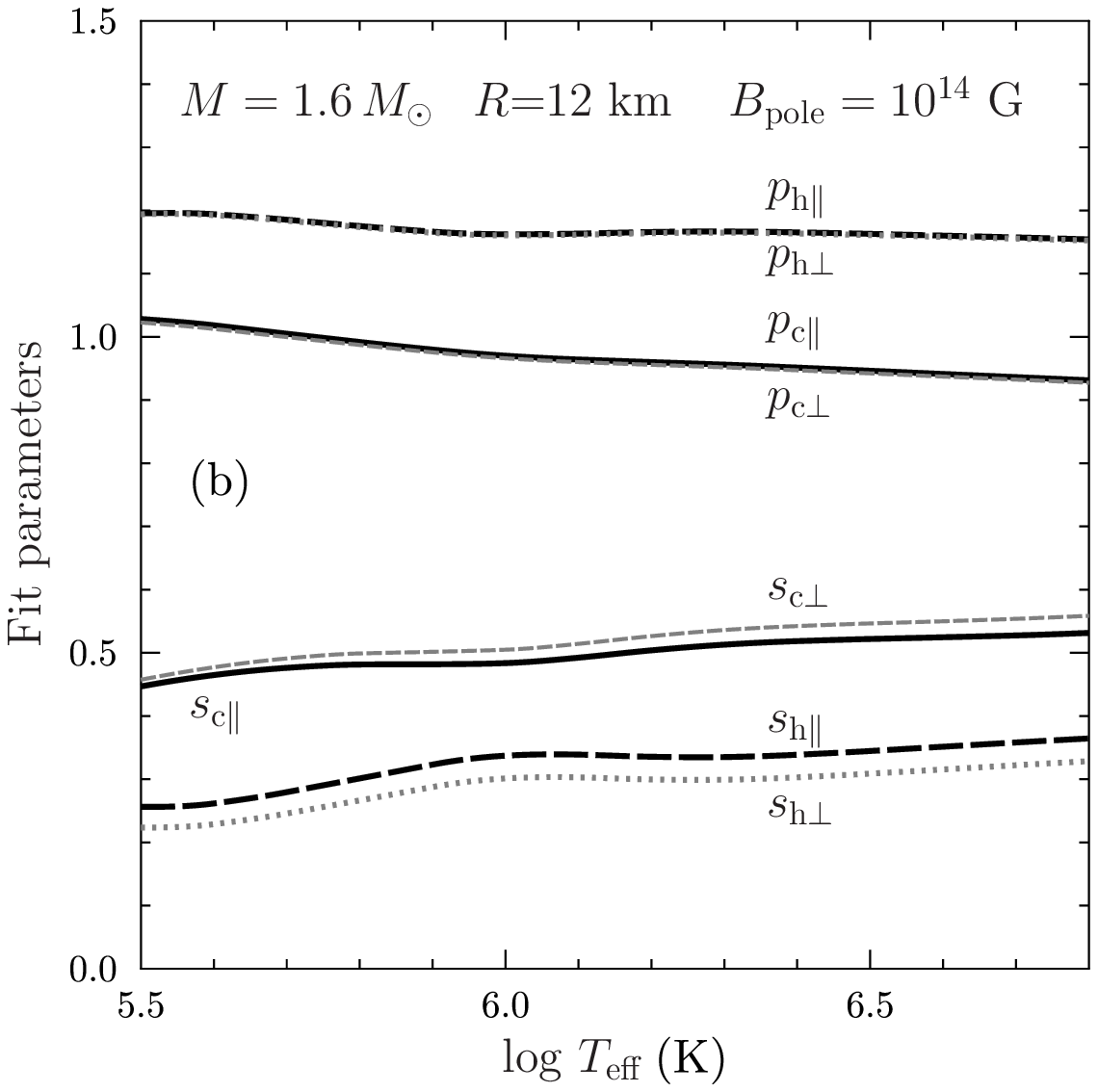}%
	\caption{
		2BB fit parameters versus  
		$\log \Teff$ for a neutron star with
		$\Bpole=10^{14}$ G (as in  
		Fig.\ \ref{f:maps1412}b at $M=1.4\,\msun$ and
		$R=12$ km) but for different $M$ and $R$.
		Panel (a) refers to a less compact star with $M=1.4\,\msun$ and
		$R=16$ km. 
		Panel (b) is for a more compact star 
		with $M=1.6\,\msun$ and 
		$R=12$ km.
	}
	\label{f:compar1}
\end{figure*}

Phase--space maps like those in Fig.~\ref{f:maps} are convenient but take
much space. We will use a more economic format, which 
allows plotting all four fit parameters on one panel.
Fig. \ref{f:maps1412} demonstrates the behavior of fit parameters (Table \ref{tab:fit}) versus $\Teff$ 
of the colder and hotter  BB components from the star
with the same $M$ and $R$ at 
$\Bpole=10^{12}$ and $10^{14}$~G [panels (a) and (b), 
respectively].  
The fits for the pole observations are plotted in black,
while those for the equator observations in grey. The upper solid and long--dashed 
black curves show the parameters $p_{\rm c \parallel}$ and $p_{\rm h \parallel}$, which specify the temperatures 
$T_{\rm effc\parallel}$ and $T_{\rm effh\parallel}$, respectively. 
The lower solid and long--dashed 
black curves show the relative surface areas $s_{\rm c \parallel}$ and $s_{\rm h \parallel}$ 
of the colder and hotter surface regions. For the equator observations, $p_{\rm c \perp}$ 
and $p_{\rm h \perp}$ 
are plotted by the short--dashed and dotted upper grey lines, 
while $s_{\rm c \perp }$ and $s_{\rm  h \perp }$ by similar lower grey lines.

For both panels in Fig.\ \ref{f:maps1412} we
have $T_{\rm effh \parallel}\approx T_{\rm effh \perp}$
and $T_{\rm effc \parallel}\approx T_{\rm effc \perp}$. This
is because the spectral fluxes are
generally close to isotropic 1BB  with given $\Teff$ (\citealt{page95}; also see Fig.\ \ref{f:basic}). However,
$T_{\rm eff h}$ is noticeably higher than $T_{\rm eff c}$, indicating that hotter and colder surface regions are
distinctly pronounced in the spectra. Also, we have $s_{\rm c \parallel} \approx  s_{\rm c \perp}$ and $s_{\rm h \parallel} \approx  s_{\rm h \perp}$. 

The dependence of fit parameters on $\Teff$ is
seen to be smooth and not too sensitive
to $\Bpole$. Systematically, $T_{\rm effh}$ is
from 10 to 20 per cent higher than $T_{\rm effc}$.
In a colder star ($\log \Teff \sim 5.5$) one has 
$\Teff \approx T_{\rm effc}$, but for a hotter star
($\log \Teff \gtrsim 6.5$)  $\Teff \approx T_{\rm effh}$. 
The depencence of the fractional areas 
$\scc$ and $\shh$ on the observation angle is stronger than for $\pcc$ and $\phh$.

Fig.\ \ref{f:mapsofB} shows the behavior of the same 2BB fit parameters, as in Figs.\ \ref{f:maps1412} 
($M=1.4\,\msun$, $R=12$ km), but versus $\Bpole$ at $\log \Teff$[K] =6.4 [panel (a)] and 5.8 [panel (b)].
The dependence of fit parameters on $\Bpole$ is weak.
For instance, $\Bpole$ varies over 3.5 orders of magnitude
in Fig.\ \ref{f:mapsofB} but the fit parameters vary much less. 

Fig.\ \ref{f:compar1} shows the same 2BB fit parameters versus $\Teff$ at $\Bpole= 10^{14}$ G, as in
Fig.\ \ref{f:maps1412}b, but for other values 
of $M$ and $R$. Fig.\ \ref{f:compar1}a refers to a less
compact star ($M=1.4\,\msun$, $R=16$ km, smaller $M/R$), 
while Fig.\ \ref{f:compar1}b is for a more
compact star, with $M=1.6\,\msun$ and $R=12$ km.
The overall  behavior of the curves in panel (b) 
is similar to that in panel (a). Generally, 
the parameters $p$ on panel (b) are less affected by the stellar compactness than the parameters $s$. The stronger
the compactness, the smaller the difference between the fit parameters for the
pole and equator observations, making the thermal radiation more uniform due to stronger
general relativisitc (GR) effects. This was shown by
\citet{page95} by direct calculations of some spectral fluxes.

Let us stress that we have fitted {\it unabsorbed}
computed spectral fluxes by {\it unabsorbed} 2BB models. In this way
we have established  one-to-one theoretical correspondence between the two models. On the other hand, observational data allow observers (in principle)	
to infer 2BB emission parameters, which are corrected for interstellar absorption, for a possible existence of non-thermal radiation component and similar
effects. If so, we can compare these observational data
with the parameters given by our unabsorbed 2BB models considered here.

So far we have studied 2BB maps for neutron stars with
heat blankets made of iron (e.g. \citealt{PYCG03,BPY21}). As a test, we have constructed a 2BB map for the star with $M=1.4\,
\msun$, $R=12$ km and $\Bpole=10^{14}$ G versus $\log \Teff$ using the blanket
composed of accreted matter. We do not present the plot; it looks similar to that
in Fig.\ \ref{f:maps1412}b.

\subsection{Phase-resolved spectroscopy}
\label{s:PhaseResolved}

Phase-resolved spectroscopy from magnetized spherical neutron stars
with dipole magnetic fields in the surface layers was extensively 
studied by \citet{page95} by calculating a number of phase-resolved
spectra.

We could do the same using our 2BB fits to the spectral fluxes for
the pole and equator observations and composing the fluxes for any observation direction (Sect.\ \ref{s:2BBfit-anyi}). The results would be similar. Some differences would stem from using somewhat more refined surface temperature distribution \citep{PCY07}
than in \cite{page95}. We do not describe the details but refer to \citet{page95} for an analysis of
pulse fractions. For a  star with $M=1.4\,\msun$, $R=12$ km, $\Bpole \sim 10^{12}$ G and $\log \Teff$[K]$\sim 5.6$ at $E\sim 3$ keV the maximum pulse fraction is about 10 per cent. It increases with
the growth of $\Bpole$, $E$, and $R/M$ but decreases with the growth of $\Teff$. It could be significantly higher if one neglected the effects of light bending above the neutron star surface.  Other models of emission from 
magnetized neutron stars including pulse resolved spectroscopy, 
more complicated field geometry and
a possible presence of condensed surface, have been
considered in many publications (e.g. \citealt{page96,perez-azorin2005,geppert2006,zane-turolla2006,
Hambaryan_ea11,popov17,grandis21}).

\subsection{Testing cold equatorial belt}
\label{s:coldbelt}

So far we have studied spectral fluxes 
and their 2BB maps for one particular model (Sect. \ref{s:Palex model}) of the
surface temperature distribution $\Ts (\vartheta)$, 
corresponding to a pure dipole magnetic field
in the surface layer of the star. 
Obviously, realistic $\Ts (\bm{n})$ distributions can be more
complicated. For simplicity, we
restrict ourselves to sufficiently weak axially symmetric variations of $\Ts(\vartheta)$, which are symmetric also with respect to the magnetic
equator. 

The most doubtful place in our $\Ts (\vartheta)$ distribution is
a thin cold equatorial surface belt (Sect.\ \ref{s:Palex model},
Fig.\ \ref{f:Rsoftheta}). The temperature is much lower there, than
in other places of the surface. Because this temperature
distribution was calculated (e.g. \citealt{BPY21}) taking into
account only the radial heat transport, the equatorial belt
is the first place, where this $\Ts (\vartheta)$ can be questioned.
The heat can spread along the surface of the cold belt to the equator,
reducing surface temperature variations there.

We have checked this effect on phenomenological level. We have
taken a narrow equatorial surface 
strip of angular half-size $\delta \vartheta_{\rm belt}$
and left the $\Ts (\vartheta)$ dependence untouched outside this strip
($|\vartheta- \vartheta_{\rm eq}| \geq \delta \vartheta_{\rm belt}$,
with $\vartheta_{\rm eq}=90^\circ$).
Inside the strip, we have artificially assumed that
$\Ts (\vartheta)= (A-
C \sin^2 \vartheta)^{1/4}$,
where constants $A$ and $C$ have been chosen to ensure continuity of
$\Ts(\vartheta)$ and 
$\partial \Ts/\partial \vartheta$ at $\vartheta=\vartheta_{\rm belt}=\delta_{\rm eq}-\delta_{\rm belt}$.
For $\delta \vartheta_{\rm belt}=30^\circ$, these profiles are shown by
dotted lines in Fig.\ \ref{f:Rsoftheta} (for the same values of $\Teff$ as the standard solid lines, but using modified  calculation of
$\Teff$ for the new profiles).

We have calculated modified spectral fluxes and 2BB maps for the
same conditions, as in Fig.\ \ref{f:maps1412}b, with two values
$\delta \vartheta_{\rm belt}=10^\circ$ and $30^\circ$.
In the first case the maps exactly coincided with the initial 
ones but the surface temperature still varied strongly over 
the belt. In the second case the $\Ts$ variations are
greatly smoothed out (Fig.\ \ref{f:Rsoftheta}) but the maps appeared 
almost the same as the initial ones. 

Therefore, variations of $\Ts$ in the cold equatorial belt seem to have almost no effect on the thermal surface emission (as has been anticipated earlier, e.g. \citealt{BPY21}).

\subsection{Warmer polar spots}
\label{s:hotspot}

\begin{figure}
	\includegraphics[width=0.4\textwidth]{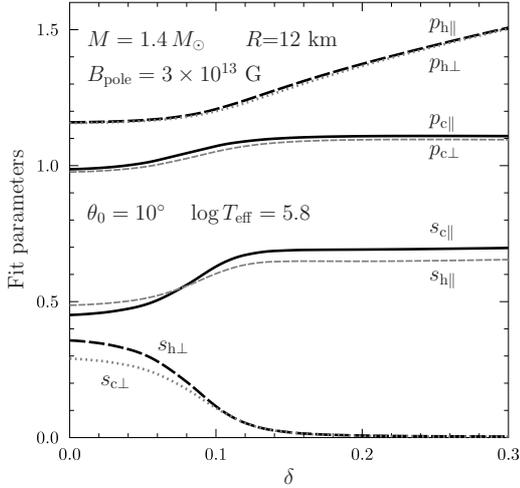}%
	\caption{
		2BB fit parameters for a neutron star with
		$M=1.4\,\msun$, $R=12$ km, $\log \Teff$[K]=5.8 and $\Bpole=3 \times 10^{13}$ G with extra hotspots of angular size
		$\vartheta_0=10^\circ$ on magnetic poles [described by
		equation (\ref{e:Ts})] versus extra relative surface
		temperature increase $\delta$ at the pole.
		See the text for details.	
	}
	\label{f:hotspots}
\end{figure}

Here we introduce two symmetric slightly warmer phenomenological spots on  magnetic poles.
Let $T_\mathrm{s0}(\vartheta)$ be the local effective surface temperature given by our
basic model (Sect.\ \ref{s:Palex model}) and $\vartheta_0$ be a small angle that deteminies the size of 
hotspots. We assume that 
\begin{equation}
   \Ts(\vartheta)=T_\mathrm{s0}(\vartheta)\,
   \left[1+\delta \, \cos^2 \left( \frac{\pi \vartheta}{2\vartheta_0} \right)\right]
   \quad \mathrm{at}~\vartheta \leq \vartheta_0,
   \label{e:Ts}
\end{equation} 
and that $\Ts(\vartheta)=\Ts(\vartheta_0)$ is not disturbed outside hotspots.  The parameter $\delta$
determines a small extra temperature enhancement in the spot's center. This enhancement 
smoothly disppears as $\vartheta \to \vartheta_0$. The presence of spots 
renormalizes (slightly increases) the total effective temperature $\Teff$,
equation (\ref{e:Ls}). 

Otherwise computations of spectral fluxes are
the same as in Sect.\ \ref{s:Palex model}.
These fluxes can also be approximated by 2BB fits 
as in Sect.\ \ref{s:2BBfits}, and 
can be analyzed via phase--space maps. For the examples presented below
the relative the fit accuracy becomes worse (reaching sometimes 10 per cent) but the fit is still 
sufficiently good and
robust.

Fig.\ \ref{f:hotspots} plots the illustrative 2BB fit parameters versus $\delta$ for a star
with 1.4\,$\msun$, $R=12$ km, $\Bpole=3 \times 10^{13}$~G,
$\log \Teff$[K]=5.8 and hotspot angular size $\vartheta_0=10^\circ$. In the absence of hotspots ($\delta=0$),
the results are very close to those in Fig.\ \ref{f:maps1412}b calculated
at the same $\log \Teff$[K]=5.8 (but for slightly higher
$\Bpole=10^{14}$~G). However, when
$\delta$ grows up, the fit parameters become different. The temperature $T_{\rm effh}$ of the hotter
BB component  and the effective 
emission surface area $s_{\rm c}$ of the colder BB component noticeably increase,
whereas the emission surface area of the hotter component dramatically
falls down. Even with really small hotspot temperature enhancement
$\delta \gtrsim 0.15$ (equivalent to 0.6 per cent luminosity enhancement), we obtain qualitatively new 
phase--space portrait with $s_{\rm h} \ll s_{\rm c}$. Such 2BB fits
have been often obtained from observations of cooling isolated neutron stars
(see Sect.\ \ref{s:discuss}); these sources are usually interpreted
as neutron stars with small hotspots. 

Therefore, the theory predicts two types of 
neutron stars whose spectra are described by the 2BB models.
The first ones are those with smooth surface temperature
distributions, created by nonuniform surface magnetic fields and
considered mainly thoughout this paper. We will call these
spectral models as 2BB {\it with smooth magnetic atmospheres}.
The sources of the second type are those with distinct hotspot
BB component. We will call them as 2BB
{\it with hotspots}. Obviously,
the sources of the two types are different but there is 
 a smooth transition between them
(for instance, by increasing $\delta$ in Fig.\ \ref{f:hotspots})
meaning possible existence of the sources of intermediate type. Such a transition would require some extra heating
in the hotter place of the magnetic atmosphere.  
According to Sect.\ \ref{s:coldbelt}, an
extra heating of the colder equatorial belt would be less efficient.

\section{Discussion and conclusions}
\label{s:discuss}

\subsection{Candidates}
\label{s:candidates}

Let us mention several candidates, among cooling isolated
middle-aged neutron stars
(not magnetars), which may demonstrate their  
multi--BB spectra as 2BB ones. We do not pretend to be complete, but give some examples. 
We base on the recent catalog \citep{Catalog20} 
available also on the Web: \text{//www.
ioffe.ru/astro/NSG/thermal/cooldat.html}.

1. XMMU J172054.5$-$372652 is a neutron star that is 
probably associated with the SNR G350.1$-$0.3 (as suggested by 
\citealt{Gaensler2008-cco}). There is no direct evidence
of pulsations. \citet{Lovchinsky_11} presented
arguments that the SNR G350.1$-$0.3 is
freely expanding and estimated its age as $\sim 1000$
years.
\citet{Catalog20} used archival \chan\ data and fitted the X-ray
spectrum with a neutron star 
(\textsc{nsx}) atmosphere model assuming
$M = 1.4 \msun$ and $R = 13$ km. They obtained
$\Teff \approx 2$ MK, but did not perform
two-component fits which would be interesting.

2. PSR B1055$-$52 (J1057$-$5226) is a well known moderately magnetized middle--aged pulsar.
Its effective magnetic field, as determined from the standard model of magnetic dipole
braking and reported in the ATNF pulsar catalog \citep{ATNF}, is
$\Beff=1.1 \times 10^{12}$~G (at the equator of an imaginary star with
$M=1.4\,\msun$ and $R=10$ km). This value does not allow one to accurately
determine $\Bpole$ but indicates that $\Bpole$ is roughly equals
a few times of $10^{12}$~G. \citet{Catalog20} presented the value 
$\kB T_\text{eff}^\infty \approx 70$ eV ($T_\text{eff}^\infty \approx 0.8$ MK). It is based on
the 2BB spectral fit by \citet{DeLuca_ea05} which included also a power-law (PL)
non-thermal radiation component. \citet{Catalog20} have corrected the results by
\citet{DeLuca_ea05} for more plausible distance estimate
to the source obtained by \citet{MignaniPK10}. The fit gives $T_\text{effh}/T_\text{effc} \sim 2.3$
and $\shh \ll \scc$. It agrees with a 2BB model containing hotspots.

3. PSR J1740+1000 has $\Beff =1.8 \times 10^{12}$~G. 
The  2BB spectral fit was done by \citet{Kargaltsev_ea12}. 
Taking the same version of the fit as selected by \citet{Catalog20}, we again
have $\kB T_\text{eff}^\infty \approx 70$ eV ($T_\text{eff}^\infty \approx 0.8$ MK),
$T_\text{effh}/T_\text{effc} \sim 2.8$ and $\shh \ll \scc$,  
with the same
conclusion as for the PSR B1055$-$52.

4. PSR B1823$-$13 (J1826$-$1334) is located in the SNR G18.0$-$00.7 
and has $\Beff=2.8 \times 10^{12}$~G.
Its X-ray emission is mostly non-thermal  \citep{PavlovKB08,Zhu_ea11} but the PL
fits suggest the presence of some thermal component. The 1BB+PL fit gives the radius
of thermally emitting region $\Reff \approx 5$ km, that is smaller than a realistic radius
of a neutron star. Adding the second BB component does not seem statistically sound
with the present data but might be possible in the future.

5. RX J1605.3+3249 (RBS 1556) is a neutron star studied  
by many authors (e.g. \citealt{Motch_05,Posselt_ea07,Tetzlaff_ea12,Pires_ea19,
Malacaria_19}) with contradictionary conclusions on its distance and
other properties (see \citealt{Catalog20}, for details). 
Its spin period has been been found but later
disproved by \citet{Pires_ea19}. The timing and spectral analysis
of X-ray emission have been performed under different assumptions using
BB and neutron star atmosphere models.
Recently \citet{Pires_ea19} have analysed the \xmm\
observations. Also, \citet{Malacaria_19}
performed a joint analysis of the \textit{NICER} and
\textit{XMM-Newton} data. Both teams report improving of
2BB fits if one adds a broad Gaussian absorption line.
In this case, they obtain $\kB T_{\rm effc}^\infty \sim 60$ eV ($T_{\rm effc}^\infty \sim 0.7$ MK),
and $T_{\rm effh}/T_{\rm effc} \sim 2$. This can be the 2BB case
with not very warm spots, where the effects of smooth magnetic
atmosphere are not negligible.

6. RX J1856.5$-$3754 is a neutron star with nearly thermal
spectrum. It was discovered by \citet*{WalterWN96}. It is slowly
rotating, with the sipn period of about 7 s; its effective magnetic
field is $\Beff \sim 1.5 \times 10^{13}$~G (although maghetic  field properties are still highly debated; e.g. \citealt{popov17,grandis21}
and references therein).
Its spectrum has been measured  in a wide range of wavelengths,
including X-rays, optical and radio. It has been interpreted with
a number of spectral models, particularly, with the model of 
thin partially ionized magnetized hydrogen atmosphere 
on top of solidified iron surface (e.g., \citealt{Ho_etal07,Potekhin14}).
Note alternative 2BB,
2BB+PL, and 3BB fits constructed by
\citet{Sartore_ea12} and \citet{Yoneyama_ea17}. 
In particular, the 2BB fits give $\kB T_{\rm effc}^\infty \sim 40$~eV 
($T_{\rm effc}^\infty \approx 0.46$~MK), 
$T_{\rm effh} /T_{\rm effc} \sim 1.6$, $R_{\rm effh} \sim 0.5 R$ and
$R_{\rm effc}\sim R$. These 2BB fits look closer to the
2BB spectral models with smooth magnetic atmosphere, than other
2BB fits described above. Introducing warmer spots simplifies
this interpretation.

In summary, if 2BB spectral fits are done, they can 
give a hint on a possible importance of localized hotspots or
smooth variations of the surface temperature. 
For refining this interpretation, it would be better to use more
advanced models of neutron star atmospheres
(e.g. \citealt{HoPC08,ZKSP21}).

\subsection{Conclusions}
\label{s:conclude} 

We have studied simple models (Sect. \ref{s:model}) 
of thermal spectra emitted from surfaces of
spherical isolated neutron stars with dipole surface magnetic fields 
$10^{11} \lesssim \Bpole \lesssim 10^{14}$~G. Such fields make the surface temperature
distribution noticeably non-uniform. The model assumes BB radiation with a
local temperature from any surface element, meaning multi--BB emission as a whole.
The model 
properly treats the GR effects as well as variations of temperature
and magnetic fields over the surface (such as colder equatorial belts and hotter magnetic poles).   

We have shown that these multi--BB radiation spectra 
are accurately fitted by 2BB models for all
observation directions by specifying eight fit parameters
(temperatures and effective emitting areas of colder
and hotter BB components for pole and equator observations) at  
any given set of values of $\Bpole$,
$\Teff$, $M$ and $R$. 
For example, Table \ref{tab:fit} presents
fit parameters for a star with $M=1.4\,\msun$ and $R=12$ km
on a grid of $\Bpole$ and $\Teff$ values. We can
easily generate such tables for any values of input parameters.
We have studied the so called 2BB portraits of neutron star
emission, which give the dependence of fit parameters on input phase--space ones 
($\Bpole$, $\Teff$, etc.). Comparing these theoretical 2BB maps with
2BB fits, inferred from the observed
spectra, gives a  method of estimating the magnetic field
properties on neutron star surface. 

Next we show (Sect.\ \ref{s:coldbelt}) that a possible extra heating of cold equatorial surface belt by heat flows from neighboring warmer surface regions does not have strong impact on theoretical 2BB maps. In constrant, even slight extra heating of magnetic poles can
strongly affect the radiation spectra but does not violate
the 2BB approximation (Sect.\ \ref{s:hotspot}). It increases 
the effective temperature $T_{\rm teffh}$ of the hotter
BB component, strongly reduces the fractional emission area $s_{\rm h}$ 
of this component, drastically changes  2BB maps and
realizes a transition to 2BB emission with hotspots.
The latter regime is quite known
from observations. We have mentioned some cooling neutron stars
which emit as 2BB with hotspots or as 2BB transiting
between hot spots and smooth magnetic atmospheres (Sect.\
\ref{s:candidates}.)

Our consideration is  restricted by studying dipole--like surface magnetic fields. Actual fields can be more
complicated and the assumption that any surface element radiates as a BB can be violated, but such effects go beyond the scope of this paper. However, the present consideration can be extended to more sophisticated field geometries and the 2BB approximation can be a proper tool for theoretical interpretations of these observations as well.

\section*{Acknowledgments}

I am grateful to the anonymous referee for constructive critical remarks. 
This research was partly supported by the grant 14.W03.31.0021 of the Ministry of Science and Higher Education of the Russian Federation
and by the  travel grants 316932 and 332666 of the Academy of Finland.

\section*{Data availability}
The data underlying this article will be shared on
reasonable request to the corresponding author.

\bibliographystyle{mnras}


\label{lastpage}

\end{document}